\documentclass[pra,preprintnumbers,amsmath,amssymb,twocolumn]{revtex4}

\usepackage{graphicx}
\usepackage{color}
\usepackage[utf8]{inputenc}
\usepackage[space]{grffile}
\usepackage[english]{babel}

\begin{document}

\newcommand{\bra}[1]{\left\langle #1 \right|}
\newcommand{\ket}[1]{\left| #1 \right\rangle}
\newcommand{\cmi}{\ {\rm cm}^{-1}}
\newcommand{\polesA}{z_{a}}
\newcommand{\polesB}{z_{b}}
\newcommand{\Res}[2]{\mathrm{Res}_{#1}{[#2]}}

\graphicspath{{./figures_spd_paper_v2.0/}}


\title{Analytic Representations of Bath Correlation Functions for Ohmic and Superohmic
Spectral Densities Using Simple Poles}

\author{Gerhard Ritschel}
\affiliation{Max Planck Institute for the Physics of Complex Systems,
N\"othnitzer Strasse 38, D-01187 Dresden, Germany}
\author{Alexander Eisfeld}\email{eisfeld@pks.mpg.de}
\affiliation{Max Planck Institute for the Physics of Complex Systems,
N\"othnitzer Strasse 38, D-01187 Dresden, Germany}

\date{\today}

\begin{abstract}
We present a scheme to express a bath correlation function (BCF) corresponding to a given spectral density (SD) as a sum of damped harmonic oscillations.
Such a representation is needed, for example, in many open quantum system approaches.
To this end we introduce a class of fit functions that enables us to model ohmic as well as superohmic behavior.
We show that these functions allow for an analytic calculation of the BCF using pole expansions of the temperature dependent hyperbolic cotangent.
We demonstrate how to use these functions to fit spectral densities exemplarily for cases encountered in the description of photosynthetic light harvesting complexes.
Finally, we compare absorption spectra obtained for different fits with exact spectra and show that it is crucial to take properly into account the behavior at small frequencies when fitting a given SD.
\end{abstract}

\maketitle

\section{Introduction\label{sec:_Introduction}}

The influence of an environment on some relevant system degrees of freedom is often treated using open quantum system approaches \cite{We08__,MaKue00__,BrPe02__}.
In these open quantum system models the environment is often modeled as an infinite number of harmonic oscillators that couple linearly to some system degrees of freedom.
It is convenient to describe the influence of the environment on the system using a spectral density (SD)~\cite{We08__,MaKue00__,BrPe02__}, which contains information about the spectrum of the environment as well as the frequency-dependent coupling.
The definition of the spectral density starting from a microscopic system-environment model is briefly reviewed in appendix~\ref{sub_hamiltonian}.
The basic quantity entering the open quantum system approaches is the so-called bath correlation function (BCF),
\begin{equation}
\alpha(t)=\frac{1}{\pi}\int_{0}^{\infty}d\omega\, J(\omega)\Bigl(\coth\bigl(\frac{\omega}{2T}\bigr)\cos(\omega t)-i\sin(\omega t)\Bigr),\label{eq:_def_bath_correlation_function}
\end{equation}
 which contains the spectral density (SD) $J(\omega)$ as well as the temperature $T$.
We set $\hbar=1$ and $k_{\rm B}=1$.
Note that sometimes $j(\omega)=J(\omega)/\omega^2$ is denoted as the SD \footnote{Our choice corresponds to the Fourier transform of the energy-gap fluctuation function \cite{Mu95__}.} (e.g.\ in Ref.~\cite{MaKue00__}), which is more convenient for the interpretation of optical properties.
We will come back to this point later in this work.

For many numerical methods (see e.g.\ Refs.~\cite{MeTa99_3365_,Kl04_2505_,IsTa05_3131_,DaPoWi12_35_,SiBr12_1815_,KrKrRo11_2166_,SuEiSt14_arxiv_}) that handle these open quantum systems, it is important that the BCF can be (at least approximately) written as a {\it finite} sum of exponentials
\begin{equation}
\label{eq:def_EXP_BCF}
\alpha(t) \approx \sum_{m=1}^Mp_m e^{i \omega_m t}, \quad \quad t\ge 0
\end{equation}
 with time-independent complex prefactors $p_m$ and complex ``frequencies'' $\omega_m=\Omega_m+i \gamma_m$ where $\Omega_m$ and $\gamma_m$ are real numbers.
 Although the BCF Eq.~(\ref{eq:_def_bath_correlation_function}) is also defined for negative times, it is usually sufficient to represent the BCF at positive times, since one can always calculate numerically the part for $t<0$ from the symmetry $\alpha(-t)=\alpha^*(t)$.

Typically, the numerical cost  to calculate properties of the open quantum system grows rapidly with the number of exponentials in Eq.~(\ref{eq:def_EXP_BCF}). 
Therefore, the number of exponentials $M$ should be as small as possible.

For a general BCF an exact decomposition with a finite number of terms typically cannot be achieved.
Nevertheless, one can use the function (\ref{eq:def_EXP_BCF}) as a fit function to the BCF trying to obtain a good fit with as few terms as possible.
To this end one can use non-linear fit routines, which are availabe in many program packages, or use methods designed particular for this task, such as filter diagonalization \cite{WaNe95_8011_, MaTa97_6756_}.
However, obtaining a direct fit of the BCF is in general not trivial, because the fitting routine might depend sensitively on the initial fit parameters (see e.g.\ also the discussion in Ref.~\cite{DaWiPo12_arxiv_}).
Furthermore, it is often not easy to judge the quality of a fit of the BCF, since for different quantities that one wants to calculate in the end different properties of the fit are important, as will be discussed below.

The Fourier transform of the BCF Eq.~(\ref{eq:_def_bath_correlation_function}) contains the same amount of information as the BCF itself and is typically a function for which fitting is more intuitive.
We have used such a procedure e.g.\ in Ref.~\cite{MoReEi12_105013_}.

However, one still needs a set of useful fit functions.
Furthermore, often one encounters the situation that an SD is given.
Using the procedure described above, one would, for a given temperature, first calculate the exact BCF, which is then either fitted directly by the sum of exponentials, or one would fit its Fourier transform instead.
Besides the fact that direct fitting of these functions is not easy this has to be done for every temperature.
It is therefore desireable to obtain the form (\ref{eq:def_EXP_BCF}) directly from the SD in a simple and transparent way.
Of course, as before, one will in general not find an exact result and one needs to approximate the SD (and the hyperbolic cotangent in Eq.~(\ref{eq:_def_bath_correlation_function})) with suitable functions (some possibilities are discussed e.g.\ in Refs.~\cite{MeTa99_3365_, Kl04_2505_, BuMaHu12_144107_, DaPoWi12_35_}).

In this work we present a class of fit functions for the SD, which allow on the one hand for an analytical calculation of the terms entering in Eq.~(\ref{eq:def_EXP_BCF}) and on the other hand for an inclusion of the relevant features of the SD in a clear way.
Our choice for the fit functions is motivated by the fact that approximating the SD by a sum of simple poles, together with a pole approximation of the hyperbolic cotangent, leads to a form of Eq.~(\ref{eq:def_EXP_BCF}).
Such pole expansions have been used in various contexts in the literature (see e.g.\ Refs.~\cite{TaKu89_101_, ChMu96_4565_, MeTa99_3365_, BaChMu02_143_, ToSm02_3848_, IsTa05_3131_, To09_094501_, DaPoWi12_35_}).
When fitting the Fourier transform of the BCF with simple poles it is (similar to the direct fitting with exponentials) usually difficult to obtain certain relevant behaviors of the SD/BCF.
Therefore, our fit functions are based on a representation $P(\omega)/Q(\omega)$, where $P(\omega)$ and $Q(\omega)$ are polynomials in $\omega$, and $P(\omega)/Q(\omega)$ can be written as a sum of simple poles.

Our approach generalizes that of Meier and Tannor \cite{MeTa99_3365_}.
Meier and Tannor parametrize an SD as a sum of anti-symmetrized Lorentzians and take a few exact poles of the hyperbolic cotangent into account, which then leads to an exponential BCF.
By anti-symmetrization  of their Lorentzian fit functions Meier and Tannor achieve a linear behavior of these functions at small frequencies. 
This is the so-called ohmic case.
The class of fit functions presented in this work allows us to go beyond the ohmic case and enables us to model different low- and high-frequency behavior of the original SD in a transparent way. 
In particular we are able to treat the important case of superohmic SDs, which are relevant, for example, for light-absorbing molecules (either from experiment \cite{WePuPr00_5825_,RaeFr07_251_} or theory \cite{DaKoKl02_031919_,OlKl10_12427_,ShReVa12_649_,VaEiAs12_224103_}) or defect tunneling in solids \cite{CaLe83_374_}.
It is well known that there are qualitative differences between ohmic and superohmic SDs for calculated open system properties (see e.g.\ Refs.~\cite{KnSmMu02_1_,KeFeRe13_7317_}). 
We will discuss this exemplarily for the case of linear absorption, where one clearly sees that a proper fitting of the low-frequency part of the SD is of great importance. 
Our fit functions allow us to intuitively find good descriptions of fits that lead to the correct behavior of the spectra.
This would be hard to achieve with fit functions that consist just of a sum of arbitrarily weighted simple poles.
In the context of  absorption spectra we will also see that for this situation $j(\omega)=J(\omega)/\omega^2$ is the relevant function for which a good fit has to be obtained.

The paper is organized as follows:
In sections \ref{sec:SD} and \ref{sec:coth} we suggest a class of fit functions for a given SD and introduce the expansion of the hyperbolic cotangent that we consider.
In section \ref{sec:eval_BCF} we state the analytical result for the BCF in the form of Eq.~(\ref{eq:def_EXP_BCF}) that we obtain from our class of fit functions together with the expansion of the hyperbolic cotangent.
In section \ref{sec:examples} we discuss the application of our method. 
After a brief consideration of the established ohmic case (section \ref{sub:ohmic}), we focus on the situation of superohmic SDs in section \ref{sub:superohmic}. 
There we show explicitly how our fit functions can be used, exemplarily  for several superohmic SDs.
By considering absorption spectra, we  demonstrate in section \ref{sub:low-frequency-behavior} the importance of having a good representation of a given SD for small frequencies.
In section~\ref{sec:conclusion} we conclude with a summary and an outlook.

Calculations and useful considerations are given in several appendices:
In appendix~\ref{sec_harm_bath_model} we briefly state the open quantum system model and we define the SD. 
We also specify the model system that we use to calculate absorption.
In appendix~\ref{sec:T0_spectra} we provide analytical results for the absorption spectrum for weak system-bath coupling and low temperature.
In appendix~\ref{sec:evaluation} we explain the basic approach that we use to evaluate analytically the integral appearing in Eq.~(\ref{eq:_def_bath_correlation_function}) using the residue theorem.
In appendix~\ref{sec:coth_approx} we describe in detail how we treat the hyperbolic cotangent with a particular focus on the structure in the complex plane.
And in appendix~\ref{sec:_SD_With_Even_Parameter_n} we discuss fit functions with even power law scalings at low frequencies, for which we could not find an exact exponential representation.

\section{Representation of the SD, the hyperbolic cotangent and the resulting BCF}

\subsection{The spectral densities}\label{sec:SD}

\begin{figure}[t]
\includegraphics[width=5cm]{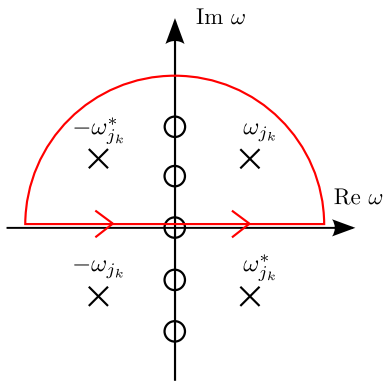}
\caption{\label{fig:skizzeInt}Sketch of the pole structure in the complex plane. The crosses denote poles of our fit functions Eq.~(\ref{eq:_sum_of_anti-symmetrized_terms_scaled}) and the circles indicate poles of the hyperbolic cotangent (or of its approximation). The depicted integration contour (red line) is used to solve the integral in Eq.~(\ref{eq:_def_bath_correlation_function}) using the residue theorem. The calculation can be found in appendix~\ref{sec:evaluation}.
}
\end{figure}

As mentioned in the introduction, our aim is to use fit functions for a given SD, that on the one hand allow us to represent the corresponding BCF as a sum of exponentials and on the other hand are in such a form that one can use them in an intuitive manner. 
To this end we consider functions of the form
\begin{equation}
J(\omega)=\omega^{n-1}\sum_{k}p_{k}(J_{k}(\omega)-J_{k}(-\omega))
\label{eq:_sum_of_anti-symmetrized_terms_scaled}
\end{equation}
with $n$ odd and  $p_{k}$ being  positive  numbers and 
\begin{equation}
J_{k}(\omega)=\prod_{j_{k}}\frac{1}{(\omega-\omega_{j_{k}})(\omega-\omega^*_{j_{k}})}
\label{eq:_pole_decomposition}
\end{equation}
with $\omega_{j_{k}}=\Omega_{j_{k}}+i\gamma_{j_{k}}$, where $\Omega_{j_{k}}$ and $\gamma_{j_{k}}$ are positive ($>0$) real numbers \footnote{We use the same symbol $J(\omega)$ to denote a given spectral density as well as our fit functions Eq.~(\ref{eq:_sum_of_anti-symmetrized_terms_scaled}), which we use to approximate the former.}.
Thus our fit functions contain only simple poles that are {\it not} located on the imaginary axis (see Fig.~\ref{fig:skizzeInt}).

The prefactor $\omega^{n-1}$  in Eq.~(\ref{eq:_sum_of_anti-symmetrized_terms_scaled}) determines the behavior of $J(\omega)$ for {\em small} frequencies. 
The product structure in Eq.~(\ref{eq:_pole_decomposition}) allows one to obtain different power law scalings at {\em high} frequencies.
Note that the summand $J_{k}(\omega)$ with the smallest number of poles determines the large-frequency behavior. 
For a given power law scaling $n$ at small frequencies, the power law scaling at large frequencies is given by $n-2\kappa-2$, where $\kappa$ denotes the number of poles $\omega_{j_k}$ of this summand lying in the upper right part of the complex plane (see Fig.~\ref{fig:skizzeInt}).
Thus we can easily adjust the small- as well as the large-frequency behavior of our fit functions.

Later it will become clear why we restrict our fit functions to odd $n$. (See in particular appendix~\ref{sec:_SD_With_Even_Parameter_n} for a discussion of the case for even $n$.)

For each pole $\omega_{j_{k}}$ we have also included the corresponding complex conjugate pole  $\omega^*_{j_{k}}$ to ensure  that for real arguments our fit functions are real, too.

\subsection{Expansions for the hyperbolic cotangent}\label{sec:coth}

The hyperbolic cotangent appearing in Eq.~(\ref{eq:_def_bath_correlation_function}) has infinitely many poles on the imaginary axis---the so-called Matsubara poles, which are located at $\nu_\ell = i \pi \ell$ with integer $\ell$.
As emphasized before, we are interested in a representation of the BCF with as few exponentials as possible.
Thus we are seeking for an appropriate approximation of the hyperbolic cotangent that can be combined with the SDs of the previous subsection.
We use
\begin{equation}
\label{eq:def_C(x)}
\coth(x)\approx\frac{1}{x}+\sum_{\ell=1}^L\eta_{\ell}\Big(\frac{1}{x-\xi_{\ell}}+\frac{1}{x-\xi^*_{\ell}}\Big)\equiv C(x)
\end{equation}
 with simple complex poles $\xi_{\ell}$ and $\xi_{\ell}^*$ with  ${\rm Im}\ \xi_{\ell}>0$ and residues $\eta_{\ell}$. $L$ denotes the number of expansion terms.
The pole at $x=0$ plays a special role and will be discussed later, see appendix~\ref{sub:evaluation_a(t)}).
In appendix~\ref{sec:coth_approx} we  discuss various possibilities to obtain such an expansion (Matsubara, Croy/Saalmann, Pad\'{e}).
It turned out \cite{HuXu10_101106_, HuLuJi11_244106_} that a  Pad\'{e} approximation has superior convergence properties compared to the other two expansions mentioned above.

Beside the form of Eq.~(\ref{eq:def_C(x)}), which is essentially an expansion around small frequencies (or high temperatures), there exist also other very good approximations of the hyperbolic cotangent, which are based on a high-frequency expansion and have a form $\coth(x)\approx 1 +  \sum^L_{\ell=1} 2 \, \exp(-2 \ell x) + r(x)$, where $r(x)$ is a function that tries to take low frequencies into account. 
A particular appealing form has been suggested by Jang et al.\ in Ref.~\cite{JaCaSi02_8313_}, where $L=2$ and $r(x)=\exp(-5x)/x$. 
It thus possesses only four simple terms and is able to approximate the hyperbolic cotangent very well on the real axis.
Unfortunately, we were only able to obtain a representation as a finite sum of exponentials according to Eq.~(\ref{eq:def_EXP_BCF}) for the case $L=0$ and $r(x)\equiv 0$, as can be seen from the calculations in appendix~\ref{sec:_SD_With_Even_Parameter_n}.
Nevertheless, in some cases (e.g.\ when one is not interested in the low-frequency part of the SD)  it might be suffcient to consider only the case $L=0$ and $r(x) \equiv 0$, which amounts to approximating the hyperbolic cotangent (for positive arguments) as $\coth(x) \approx 1$.

\subsection{Analytical representation of the BCF for odd $n$}\label{sec:eval_BCF}

For the above defined functions Eq.~(\ref{eq:_sum_of_anti-symmetrized_terms_scaled}) together with the approximation Eq.~(\ref{eq:def_C(x)}) one can analytically obtain the BCF as a sum of exponentials (see appendix~\ref{sec:evaluation}), which we write as
\begin{equation}
\begin{split}
\label{eq:alpha(t)_EXP_final}
\alpha(t) =
  & i \sum_k\sum_{j_k}\big( p_{k_j}^+
      e^{i \omega_{j_{k}} t}
    + p_{k_j}^-  e^{-i \omega_{j_{k}}^* t} \big)
  \\
  & + i \sum_{\ell=1}^L J(w_\ell) \eta_{\ell} e^{i w_{\ell} t} 
\end{split}
\end{equation}
with
$\omega_{\ell}=2 T \xi_{\ell}$ 
and
\begin{align}
 p_{k_j}^+=& p_k\, \omega_{j_k}^{n-1} R_{j_k} \big( \coth\big(\frac{\omega_{j_k}}{2T}\big) - 1 \big), \\
 p_{k_j}^-=&-p_k\, (\omega_{j_k}^*)^{n-1} R_{j_k}^* \big( \coth\big(\frac{-\omega_{j_k}^*}{2T}\big) - 1 \big).
\end{align}
Here,
\begin{equation}
R_{j_k}=\frac{1}{(\omega_{j_{k}}-\omega^*_{j_{k}})}\prod_{j\ne j_{k}}\frac{1}{(\omega_{j_k}-\omega_{j})(\omega_{j_k}-\omega^{*}_{j})},
\end{equation}
where the product contains all the poles of $J_k$ except $\omega_{j_k}$.

\section{Examples}\label{sec:examples}

In this section we discuss exemplarily for some SDs how one can use the formalism described above.

First we discuss ohmic SDs often used in the literature.
Then we treat selected superohmic SDs.

Since the ohmic case has been discussed in the literature to quite some extent, we will not give much details for this case but refer to the literature instead.

For the case of the superohmic SDs, we demonstrate how our fit functions can be used.
For all SDs considered, we discuss fits to the SD and the quality of the resulting BCF.
In addition we also compare absorption spectra obtained from the fit with exact ones.
This illustrates that on the level of the BCF it is difficult to judge whether a fit is suitable for the quantity one is actually interested in.
Here one also sees the strength of the fit functions suggested by us.

\subsection{Ohmic case ($n=1$)}
\label{sub:ohmic}

For $n=1$ the fit functions Eq.~(\ref{eq:_sum_of_anti-symmetrized_terms_scaled}) have a linear dependence around $\omega=0$.

\paragraph{Simple (antisymmetrized) Lorentzian:  }
A form of $J_k(\omega)$ which has been particularly often used (for example in Ref.~\cite{MeTa99_3365_}) is that of a simple Lorentzian
\begin{equation}
J_{k}(\omega) = \frac{1}{(\omega-\Omega_{k})^{2}+\gamma_{k}^{2}}.\label{eq:_lorentzian}
\end{equation}
The resulting SD $J(\omega)$ is
\begin{equation}
J(\omega)= \sum_k p_k \frac{4 \Omega \omega}{(\omega^2-\Omega^2)^2+2 (\omega^2+\Omega^2)\gamma^2+\gamma^4},
\label{eq:sum_lorentzian}
\end{equation}
which falls off like $1/\omega^{3}$ for large frequencies.

\paragraph{Drude-Lorentz spectral density:}
The Drude-Lorentz SD is given by
\begin{equation}
J_{\rm DL}(\omega)= 2\pi \lambda \omega \frac{\gamma}{\omega^2+\gamma^2},
\end{equation}
where $\lambda$ and $\gamma$ are real parameters.
One can write $J_{\rm DL}(\omega)=2\pi \lambda \omega \gamma(\frac{1 }{(\omega-i \gamma)(\omega+i \gamma)} )$. Thus, it seems at first sight that it is of the form of our SDs.
However, it is not since it cannot be brought into the form of Eq.~(\ref{eq:_sum_of_anti-symmetrized_terms_scaled}).
We have previously used  anti-symmetrized Lorentzians Eq.~(\ref{eq:sum_lorentzian}) to approximate this Drude-Lorentz SD \cite{RiRoSt11_113034_}.

\paragraph{Exponential cutoff}
Another often used SD is the ohmic SD with exponential cutoff
\begin{equation}
J_{\rm ohm}(\omega)=\eta \omega e^{-\omega/\Lambda},
\end{equation}
where $\Lambda$ is the cutoff frequency and $\eta$ scales the overall strength.
In Ref.~\cite{MeTa99_3365_} it has been shown that one can approximate it very well with three (antisymmetrized) Lorentzians of the form of Eq.~(\ref{eq:sum_lorentzian}).

\subsection{Superohmic case}
\label{sub:superohmic}

We first turn our attention to the case where the spectral density shows a cubic behavior at frequency zero.
Then we discuss a case where the frequency dependence at zero can not be described by $\omega^n$ with odd $n$.

This section also serves to demonstrate that it is in general quite problematic to judge whether a fit of the BCF is good or not for a specific quantity that one wants to calculate for an open quantum system.
To this end we consider absorption spectra. 
The model Hamiltonian, which is briefly reviewed in appendix~\ref{sub_hamiltonian}, and the relation between the BCF and the absorption spectrum  can  be found in many publications (e.g.\ \cite{MaKue00__,Mu95__}). 
We will see, that in the presented examples a good fit to $j(\omega)=J(\omega)/\omega^2$ is the relevant quantity.

In the following we will show for each situation the SD $J(\omega)$ as well as $j(\omega)$ together with our fit.
We will also show the resulting BCFs (exact and our approximations) for three temperatures.
Furthermore, we present the absorption spectra corresponding to these BCFs.

In particular we consider three different SDs, which will be discussed in detail in the following subsections.
These SDs are shown in the first row of Fig.~\ref{fig:examples_n=5}~(a), (b), (c).
Figure~\ref{fig:examples_n=5}~(a) represents a damped molecular vibration, discussed in subsection \ref{sec:dampedMolVib}.
In  Fig.~\ref{fig:examples_n=5}~(b) the log-normal SD is shown, which has been suggested to describe the broad background observed in SDs of photosynthetic complexes, on top of which the damped vibrations sit \cite{KeFeRe13_7317_}.
This SD is discussed in detail in subsection \ref{sec:log-normal}.
Finally, in Fig.~\ref{fig:examples_n=5}~(c) the sum of the two previous SDs is shown, i.e.\ a damped molecular vibration sitting on a broad background.
Such a situation, with more molecular vibrations is typical for the SD of light harvesting systems \cite{WePuPr00_5825_,RaeFr07_251_,AdRe06_2778_}.

In this Fig.~\ref{fig:examples_n=5} we show in the second row (i.e.\ panels (d)-(f)) $j(\omega)=J(\omega)/\omega^2$ , the quantitiy that will be of particular importance for the absorption spectra.
In the third to the fifth row we show the BCFs at different temperatures, ranging from $T=4\textrm{ K}$ in panels (g)-(i), to $T=77\textrm{ K}$ in panels (j)-(l) and $T=300\textrm{ K}$ in panels (m)-(o).

Absorption spectra corresponding to the background SD and the SD of the damped vibration are presented in  Figs.~\ref{fig:spec_KJ} and~\ref{fig:spec_RE}.

A brief description of our model system, which essentially is a two-level molecule, is given in appendix~\ref{sec_harm_bath_model}.
For such a model the absorption spectrum $A(\omega)$ is given by \cite{MaKue00__}
\begin{equation}
\label{eq:abs_FT}
A(\omega)=\frac{1}{\pi}\, {\rm Re} \int_0^{\infty} dt\, e^{i(\omega-\omega_0) t} \, e^{-g(t)},
\end{equation}
where $\omega_0$ is related to the transition energy between the two states and the so-called lineshape function $g(t)$ is given by
\begin{equation}
\label{eq:def_g(t)}
g(t)=\int_0^t dt' \int_0^{t'} dt''\, \alpha(t'')
\end{equation}
and therefore depends only on the BCF.
With the relation Eq.~(\ref{eq:_def_bath_correlation_function})  between the BCF an the SD we can also write
\begin{equation}
\begin{split}
\label{eq:g(t)_SD}
g(t) = &\frac{-1}{\pi}\int_0^{\infty}\!\!\! d\omega\, \frac{J(\omega)}{\omega^2}
\Big(
\coth\frac{\omega}{2T}\big(\cos \omega t -1\big) -i \sin \omega t 
 \Big)\\
&-i E_\lambda t
\end{split}
\end{equation}
with the reorganization energy $E_\lambda=\frac{1}{\pi} \int_0^{\infty} d\omega\, J(\omega)/\omega$.
The reorganization energy $E_\lambda$ simply leads to a shift of the whole spectrum and is conveniently combined with $\omega_0$.
For all spectra we show in the following, we set the zero of the frequency axis to the position of the zero phonon line (which can formally be achieved by $\omega_0 + E_{\lambda} \equiv 0$). 
All presented absorption spectra are normalized to an area of $2\pi$.
From Eq.~(\ref{eq:g(t)_SD}) one can expect that $j(\omega)=J(\omega)/\omega^2$ is indeed the relevant quantitiy for the calculation of absorption.
In fact, in the limit of small overall coupling strength/reorganization energy and zero temperature the absorption spectrum is given by a delta peak (zero phonon line) and a phonon wing that has the shape of $J(\omega)/\omega^2$ (see appendix~\ref{sec:T0_spectra}).

Note that for our fit functions one can easily calculate the lineshape function $g(t)$ analytically, while in general it involves integration over oscillating functions \footnote{For the superohmic case $n>2$ one can solve the integral efficiently using Fourier transformation methods.}.

In all panels of the figures the exact SDs/BCFs are shown together with the results of our fit.

In Table~\ref{tab:nmb_fit_terms} we list the number of fit terms for the SDs as well as for the hyperbolic cotangent for all the fits we consider in this work. Additionally, we list the number of resulting exponential terms in the BCFs that we obtain from our fits for the different considered temperatures. The explicit fit parameters can be found in the Supporting Information~\cite{RiEi14_SI_}.

\subsubsection{Damped molecular vibration}
\label{sec:dampedMolVib}

\begin{figure}[tbp]
\includegraphics[width=3.5cm]{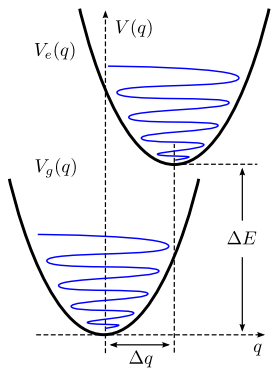}
\caption{\label{fig:VSM_sketch}Sketch of a damped molecular vibration. In both potential surfaces the vibration relaxes towards the potential minimum.  The damping is caused by the coupling of the coordinate $q$ of the mode to an environment. The potential surfaces are given by $V_g (q) = \frac{1}{2} \Omega^2 q^2$ and $V_e (q) = \frac{1}{2} \Omega^2 (q - \Delta q)^2 + \Delta E$.}
\end{figure}

\begin{figure*}[tbp]
\includegraphics[width=\textwidth]{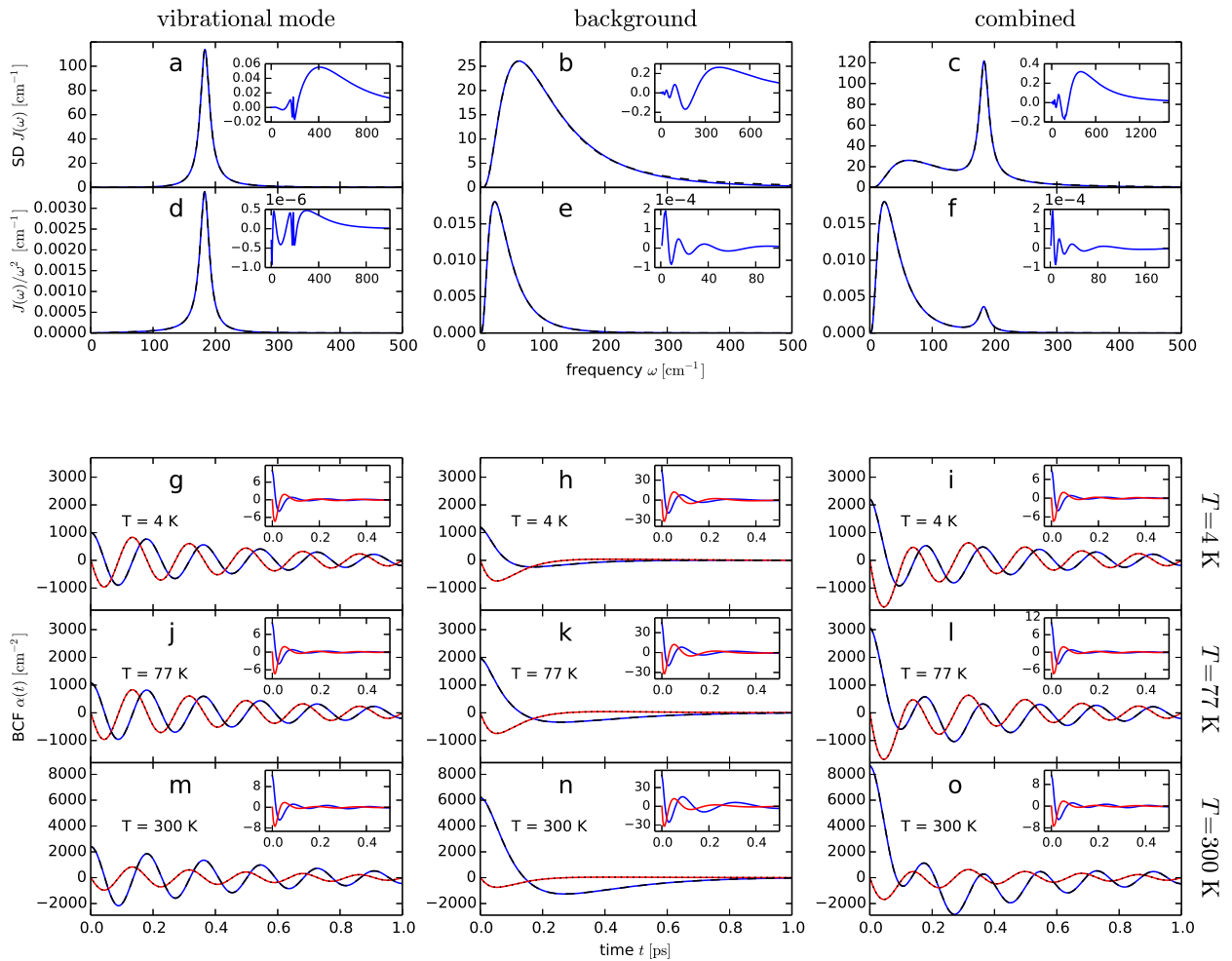}
\caption{\label{fig:examples_n=5}
The three SDs discussed in section \ref{sub:superohmic}. Left column: SD corresponding to the damped mode, Eq.~(\ref{J_ESM_Fano_Ohmic_exp_cutoff}); middle column: log-normal SD, Eq.~(\ref{eq:log-normal}); right column: sum of damped mode and log-normal SDs.
The first row (panels~(a)-(c)) shows the SD  (blue solid line) together with a fit according to the parameters of table~\ref{tab:fit_params} (black, dashed). The insets show the difference between the original SDs and the approximations.
Second row (panels~(d)-(f)): SD divided by $\omega^2$.
Rows three to five (panels~(g)-(o)): Corresponding BCFs for the three temperatures $T \in \{4\textrm{ K}, 77\textrm{ K}, 300\textrm{ K}\}$.
Blue and red correspond to the exact real and imaginary parts, respectively.
The gray dashed/dotted curves are the approximations obtained using the Pad\'e approximation of the hyperbolic cotangent.
In the insets the differences are shown.
For details see main text.
}
\end{figure*}

The first SD we consider is that of a vibrational mode related to shifted harmonic potential surfaces (with frequency $\Omega$  and Huang-Rhys factor $X=(\Omega\Delta q)^2/2$, where $\Delta q$ is the shift of the surfaces, see Fig.~\ref{fig:VSM_sketch}) that is coupled to an ohmic bath with exponential cutoff $J_{\rm ohm}(\omega) = \eta \omega e^{-\omega/\Lambda}$.
This model is discussed e.g.\ in  Ref.~\cite{RoStWh12_204110_}, where it is shown that it leads to a spectral density of the form
\begin{align}
\label{J_ESM_Fano_Ohmic_exp_cutoff}
  &J_{\rm vib}(\omega)=
\frac{X\ \omega^2 J_{\rm ohm}(\omega)} {\left(\omega - \tilde{g}(\omega))^2 + J^2_{\rm ohm}(\omega)\right)}
\end{align}
with
 $\tilde{g}(\omega) = \Omega - {\eta}\Lambda/\pi + J_{\rm ohm}(\omega) {\operatorname{Ei}(\omega/\Lambda)}/{\pi}$ and $\operatorname{Ei}(z)=-\int_{-z}^{\infty}dt\,\frac{e^{-t}}{t}$. 
This SD has a cubic behavior at small $\omega$ and falls off like $e^{-\omega}/\omega$ for large $\omega$.

One example of such an SD is shown in Fig.~\ref{fig:examples_n=5}~(a)  as solid blue line.
The used parameters are $\eta= 0.3$, $\Lambda= 100\cmi$, $\Omega=180\cmi$ and  $X=0.03$.
They correspond roughly to a low energy vibrational mode typical for bacteriochlorophyll molecules \cite{WePuPr00_5825_,AdRe06_2778_,RaeFr07_251_,KeFeRe13_7317_}.
In addition to the original SD we show a fit with a function of the form of our Eq.~(\ref{eq:_sum_of_anti-symmetrized_terms_scaled}) (dashed, black).
The parameters used are given in Table~\ref{tab:fit_params}.
The difference between the original function and our fit is shown in the inset.
One sees that already with three poles one gets a very good agreement.

As discussed in the introduction, often a different convention for the SD is used which corresponds to $j(\omega) = J(\omega)/\omega^2$ in our notation.
We show this in the second row (panel~(d)).

In the rows 3-5 (panels (g), (j), (m)) we finally show the corresponding BCFs at three different temperatures.
The imaginary part is plotted in red and the real part in blue.
With dashed/dotted lines the exponential fit obtained from Eq.~(\ref{eq:alpha(t)_EXP_final}) using the Pad{\'e} approximation is displayed.
For the case of 4~K we used eleven Pad{\'e} terms. 
For 77~K two and for 300~K one Pad{\'e} term was used. 
For all temperatures the exact curves and the exponential approximation are in perfect agreement.
\begin{table}[b]
\begin{tabular}{|c||c|c||c||c|}
\hline
 &\ damped mode\ \ &\ Log-normal\ \ & &\ \ combined\ \ \ \\
\hline
\hline
$n$ &  5 & 5 & $n$ & 5  \\
\hline
$p_1$ & 1.27$\cdot 10^{4}$ &  2.42$\cdot 10^{5}$ & $p_1$ & 1.27$\cdot 10^{4}$  \\ 
$p_2$ & ---                &  ---                & $p_2$ & 2.42$\cdot 10^{5}$  \\ 
\hline
$\omega_{1_1}$ & 183+9.17 i  & 5.52+11.3 i & $\omega_{1_1}$ & 183+9.17 i   \\
$\omega_{1_2}$ & 67.6+178 i  & 15.4+34.4 i & $\omega_{1_2}$ & 67.6+178 i   \\
$\omega_{1_3}$ & 1.76+11.1 i & 51.7+102 i  & $\omega_{1_3}$ & 1.76+11.1 i  \\ \cline{4-5}
               & ---         & ---         & $\omega_{2_1}$ & 5.52+11.3 i  \\
               & ---         & ---         & $\omega_{2_2}$ & 15.4+34.4 i  \\
               & ---         & ---         & $\omega_{2_3}$ & 51.7+102 i  \\
\hline
\end{tabular}
\caption{\label{tab:fit_params}Parameters used to fit the three SDs shown in Fig.~\ref{fig:examples_n=5} (panels~(a), (b), (c)). The resulting parameters for the BCFs shown in Fig.~\ref{fig:examples_n=5} can be found in the Supporting Information~\cite{RiEi14_SI_}.
}
\end{table}

\subsubsection{Log-normal}
\label{sec:log-normal}

Another important case is a spectral density of log-normal form
\begin{equation}\label{eq:log-normal}
J_{\rm bg}(\omega)=\frac{\pi S \omega}{\sqrt{2\pi}\sigma} e^{-[\ln(\omega/\omega_{\rm c}]^2/2\sigma^2}.
\end{equation}
Such a spectral density has been suggested to describe the broad background obtained when extracting experimentally the spectral densities of bacteriochlorophyll molecules in pigment-protein complexes~\cite{KeFeRe13_7317_}.
The suggestion for the so-called Fenna-Matthews-Olson complex is shown in Fig.~\ref{fig:examples_n=5}~(b)  as solid blue line. 
The numerical values for the parameters (taken from Ref.~\cite{KeFeRe13_7317_}) are $S=0.3$, $\sigma=0.7$, $\omega_c=38\cmi$.
In addition we show a fit with a function of the form of our Eq.~(\ref{eq:_sum_of_anti-symmetrized_terms_scaled}) (dashed, black).
The parameters used for the fit are given in Table \ref{tab:fit_params}.
The quality of the fit is again very good.

Note that SDs given by Eq.~(\ref{eq:log-normal}) have a non-algebraic superohmic behavior at low frequencies.
This we cannot describe very well with a small number of poles.
Nevertheless, when looking at the corresponding BCFs (panels~(h), (k), (n)) we see that the agreement between exact calculations and our approximation is again nearly perfect.
We used fifteen Padé terms for the low temperature of 4~K, three terms for 77~K and one term for 300~K.

\subsubsection{Background plus damped vibration}

Typically the SD of pigments in an (protein) environment consists of a broad background combined with peaks belonging to molecular vibrations.
An example is shown in Fig.~\ref{fig:examples_n=5}~(c).
Here, the used spectral density is simply the sum $J(\omega)=J_{\rm vib}(\omega)+J_{\rm bg}(\omega)$, where $J_{\rm vib}(\omega)$ and $J_{\rm bg}(\omega)$ are the SDs from Fig.~\ref{fig:examples_n=5}~(a) and (b), respectively.
We approximate the resulting spectral density by the sum of the fit functions used in Fig.~\ref{fig:examples_n=5}~(a) and Fig.~\ref{fig:examples_n=5}~(b). 
Also here the quality of the fit is very good. One can achieve a similar level of accuracy for the fit through a direct fit of the combined SD with only 5 terms (instead of 6).
The corresponding BCFs at the three different temperatures are shown in panels~(i), (l), (o).
We used thirteen Padé terms for the low temperature of 4~K, three terms for 77~K and one term for 300~K.

\subsection{Importance of the low-frequency behavior}
\label{sub:low-frequency-behavior}

\begin{figure*}[tbp]
\includegraphics[width=\textwidth]{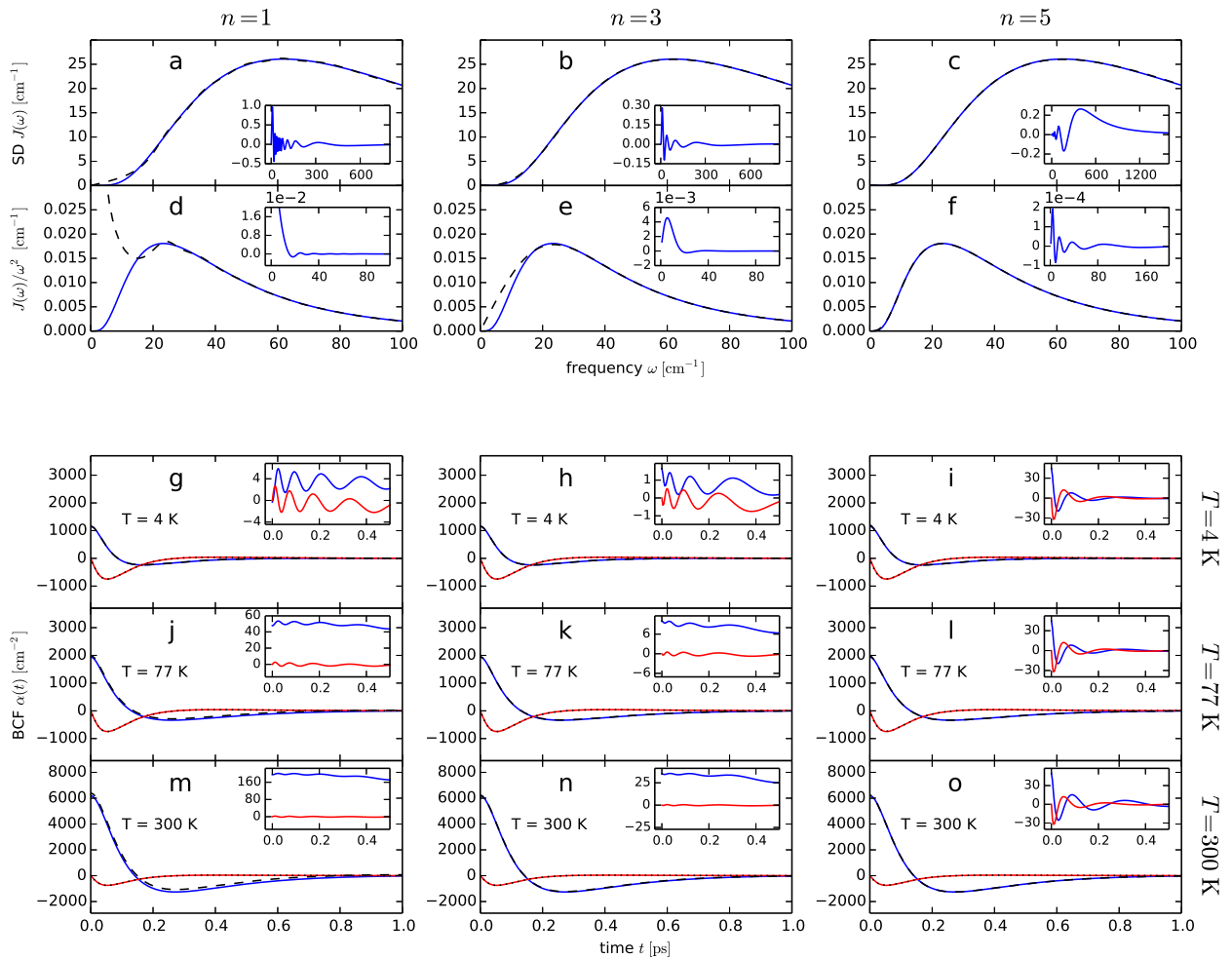}
\caption{\label{fig:n-dep}
Log-normal SD, Eq.~(\ref{eq:log-normal}) and fits for different scaling parameters $n$ in Eq.~(\ref{eq:_sum_of_anti-symmetrized_terms_scaled});
left column $n=1$, middle column $n=3$ and right column $n=5$: 
The first row (panels~(a)-(c)) shows the SD (blue solid line) together with a fit according to the parameters given in the Supporting Information (black, dashed)~\cite{RiEi14_SI_}. The insets show the difference between the original SD and the approximations.
Second row (panels~(d)-(f)): SD divided by $\omega^2$.
Rows three to five (panels~(g)-(o)): Corresponding BCFs for the three temperatures $T \in \{4\textrm{ K}, 77\textrm{ K}, 300\textrm{ K}\}$.
Blue and red correspond to the exact real and imaginary parts, respectively.
The gray dashed/dotted curves are the approximations obtained using the Pad\'e approximation of the hyperbolic cotangent.
In the insets the differences are shown.
For details see main text.
}
\end{figure*}

\begin{figure*}[tbp]
\includegraphics[width=\textwidth]{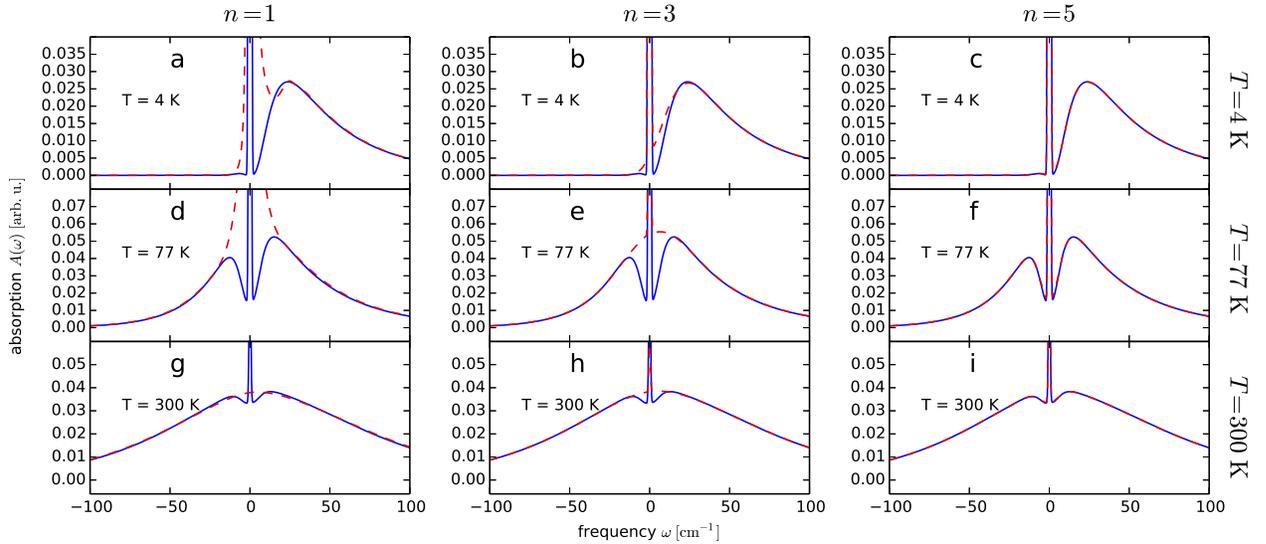}
\caption{\label{fig:spec_KJ}
Absorption spectra for the log-normal SD according to Eq.~(\ref{eq:log-normal}), which is shown in Fig.~\ref{fig:examples_n=5}~(b). The solid blue curve in each subplot is the exact spectrum and the dashed red curve the one obtained from our fit.
In the left column (panels~(a), (d), (g)) the fit was performed with $n=1$, in the middle column (panels~(b), (e), (h)) with $n=3$ and in the right column (panels~(c), (f), (i)) with $n=5$.  
The parameters for the respective fits are given in the Supporting Information~\cite{RiEi14_SI_}.
The first row is for 4~K (panels~(a)-(c)), the second for 77~K (panels~(d)-(f)) and the third for 300~K (panels~(g)-(i)).
}
\end{figure*}

\begin{figure*}[tbp]
\includegraphics[width=\textwidth]{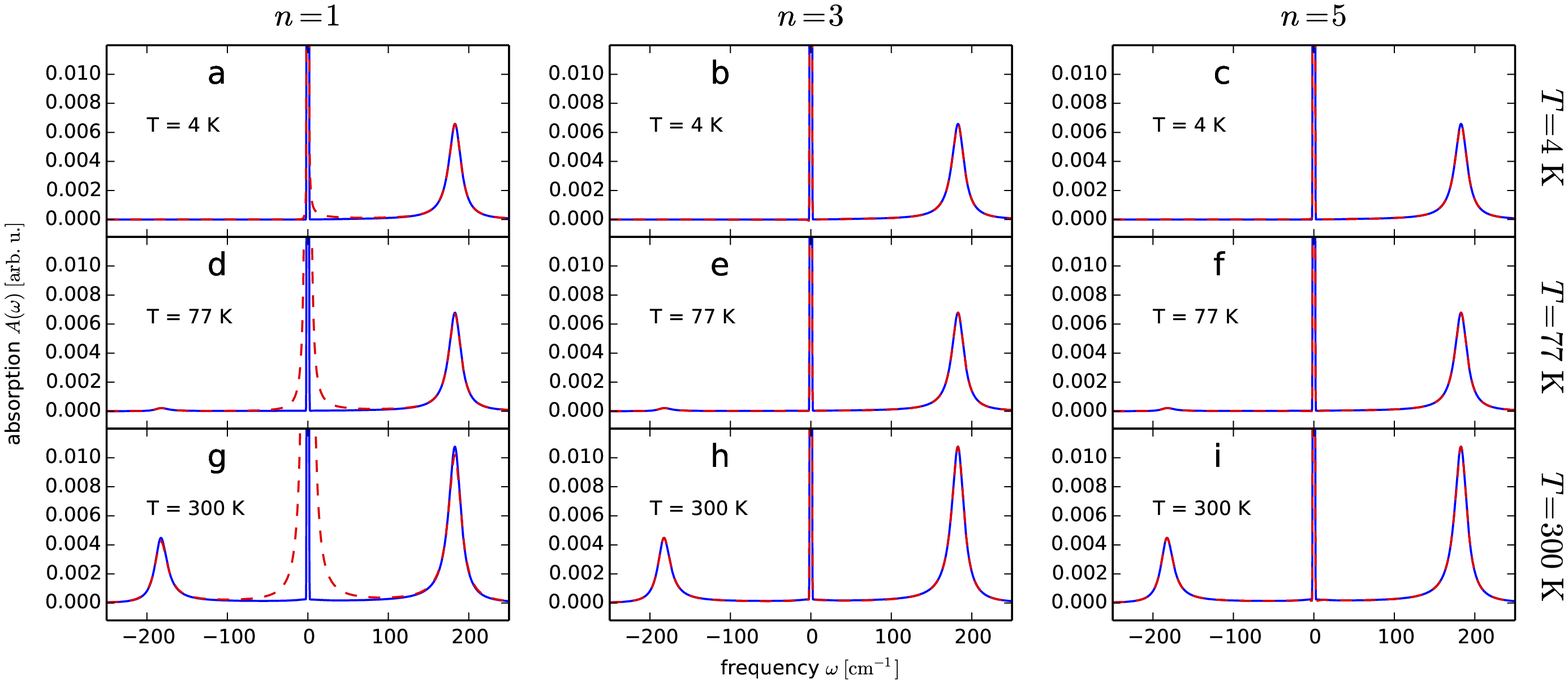}
\caption{\label{fig:spec_RE}
Same as Fig.~\ref{fig:spec_KJ}, but now for the SD of the ``damped vibration'' Eq.~(\ref{J_ESM_Fano_Ohmic_exp_cutoff}), which is shown in Fig.~\ref{fig:examples_n=5}~(a). The parameters used to fit the SD are given in the Supporting Information~\cite{RiEi14_SI_}.
}
\end{figure*}

In this section we discuss the relevance of a good fit of the low-frequency behavior  of a given SD for the  calculation of correct absorption spectra.
To this end we focus on the log-normal SD of the previous section.
In Fig.~\ref{fig:n-dep}~(a)-(c) this SD is shown again (blue solid line) together with different fits (black dashed lines). 
In the first column (i.e.\ Fig.~\ref{fig:n-dep}~(a)) the fit was performed using ohmic functions of the Meier/Tannor type (i.e.\ $n=1$).
In the middle column (Fig.~\ref{fig:n-dep}~(b)) the fit is done for $n=3$ and in the right column (Fig.~\ref{fig:n-dep}~(c)) the case $n=5$ is shown.
For $n=3$ and $n=5$ we take products with 3 poles $\omega_{j_k}$ for each summand in Eq.~(\ref{eq:_sum_of_anti-symmetrized_terms_scaled}) into account in order to adjust the high-frequency behavior of our fit functions. For $n=1$ we take 2 poles for every summand. For the cases $n=3$ and $n=5$ only one summand is necessary to obtain the fits that are shown, whereas for $n=1$ we use 4 summands (which amounts in total to 8 poles $\omega_{j_k}$ in the upper right part of the complex plane for $n=1$) in order to enhance the agreement between the SD and our fit for smaller frequencies. The resulting fit parameters are given in the Supporting Information~\cite{RiEi14_SI_}.

In Fig.~\ref{fig:n-dep} one sees that for $J(\omega)$ (first row, panels (a)-(c)) all fits look 'reasonable', although for $n=1$ (panel~(a)) there are deviations at low frequencies visible.
However, when considering $J(\omega)/\omega^2$ (second row, panels (d)-(f)) one clearly sees that there are quite large deviations at small frequencies for $n=1$ and $n=3$.
In the corresponding BCFs shown in (Fig.~\ref{fig:n-dep}~(g)-(o)), however, the huge deviations between the approximations and the exact results are hard to recognize, in particular for the low temperature case $T=4\textrm{ K}$ (panels (g)-(i)).
For $n=1$ we used eleven Padé terms in the expansion for the hyperbolic cotangent for 4~K, two terms for 77~K and one term for 300~K. For $n=3$ we used fourteen, two and one term, respectively and for $n=5$ we used fifteen, three and one term.

When considering the corresponding absorption spectra (see Fig.~\ref{fig:spec_KJ}),  one notices huge deviations for the case $n=1$ (panels (a), (d), (g)). 
 But also for $n=3$ there are considerable deviations (in particular for $T=77\textrm{ K}$, panel~(e)).

The important point to note is that it is necessary to properly approximate the low-frequency behavior of $j(\omega)=J(\omega)/\omega^2$ in order to obtain correct absorption spectra in the region of small frequency. 
Since for the log-normal SD $j(\omega)$ starts off at frequency zero with vanishing slope, fits with $n=1$ or $n=3$ are not adequate.
For $n=1$ every fit diverges at frequency zero when looking at $j(\omega)$, and for $n=3$ one always obtains a linear slope.

One might think that for a SD like that of Fig.~\ref{fig:examples_n=5}~(a), where one has a spike at relatively large frequencies and only a slow rise of the SD at low frequences, it is not so important to use a sufficiently large $n$. 
That this conjecture is not correct is demonstrated in Fig.~\ref{fig:spec_RE}, where the exact absorption spectra for the SD of Fig.~\ref{fig:examples_n=5}~(a) are plotted (solid, blue) together with the ones obtained from fits with different power laws at small frequencies (red, dashed).
This SD has a low-frequency behavior proportional to $\omega^3$, thus we expect that by using fit functions with $n=3$ we can obtain very good results. 
This is indeed the case for all temperatures considered (see Fig.~\ref{fig:spec_RE}, panels~(b), (e), (h)).   
Also with a larger power $n=5$ the spectra are well reproduced (see Fig.~\ref{fig:spec_RE}, panels~(c), (f), (i)).
However, for $n=1$ there are appreciable deviations from the exact spectrum (see Fig.~\ref{fig:spec_RE}, panels~(a), (d), (g)).
These deviations become larger for increasing temperatures.

We would like to point out that in order to enhance the agreement of our fit of the log-normal SD for $n=1$ we already used more summands than for the cases  $n=3$ and $n=5$ (8 instead of 3) in Eq.~(\ref{eq:_sum_of_anti-symmetrized_terms_scaled}) (see Table~\ref{tab:nmb_fit_terms}). 
This allowed us to reduce the error of our fit for the ohmic fit functions to some extent. 
One could, of course, use even more ohmic terms to further narrow the region of small frequencies, where the fit is bad.
However, it is not possible to remove the divergence at $\omega=0$ that one obtains for $J(\omega)/\omega^2$ with a finite number of ohmic fit terms. 
Additionally, one also has to pay the price of having a larger number of exponential terms for the BCF in the end.
Using our fit functions with larger $n$ therefore not only leads to a better approximation of the BCF in those cases (considering absorption spectra), but can also provide an exponential decomposition of the BCF with less terms than for ohmic fit functions.

Finally, we also would like to note that in order to correctly obtain the absorption spectra one also needs more Padé expansion terms than one might have expected from looking only at the differences of the BCFs.
If one does not use enough expansion terms the absorption spectra can show unphysical negative contributions, although the BCFs look perfectly fine.
This again stresses the fact that one should not judge the quality of a fit to a given SD together with an expansion for the hyperbolic cotangent solely by looking at the BCF.

\begin{table}[bt]
\begin{tabular}{l|p{6mm}p{6mm}p{6mm}|p{6mm}p{6mm}p{6mm}|p{6mm}p{6mm}p{6mm}}
       SD          & \multicolumn{3}{c|}{damped mode}                & \multicolumn{3}{c|}{log-normal}                 & \multicolumn{3}{c}{combined}                    \\ \hline
$n$                &\ \ $\mathbf{1}$ &\ $\mathbf{3}$ &\ $\mathbf{5}$ &\ \ $\mathbf{1}$ &\ $\mathbf{3}$ &\ $\mathbf{5}$ &\ \ $\mathbf{1}$ &\ $\mathbf{3}$ &\ $\mathbf{5}$ \\ \hline 
$\kappa$           &\ \  2           &\ 2            &\ 3            &\ \ 8            &\ 3            &\ 3            &\  10            &\ 5            &\ 6            \\ \hline
$T=4\textrm{ K}$   &\ \  8           & 10            & 11            &\  11            & 14            & 15            &\ \ 9            & 14            & 13            \\
$T=77\textrm{ K}$  &\ \  1           &\ 2            &\ 2            &\ \ 2            &\ 2            &\ 3            &\ \ 2            &\ 2            &\ 3            \\
$T=300\textrm{ K}$ &\ \  1           &\ 1            &\ 1            &\ \ 1            &\ 1            &\ 1            &\ \ 1            &\ 1            &\ 1         
\end{tabular}
\caption{\label{tab:nmb_fit_terms}
Number of fit terms used for the different fits presented in this work.
Second row: scaling parameters $n \in \{1, 3, 5\}$ of the different fits Eq.~(\ref{eq:_sum_of_anti-symmetrized_terms_scaled}); 
Third row: number of poles in the upper right part of the complex plane $\kappa$ that we used to approximate the SDs for the different $n$. 
In the last three rows we list the number of expansion terms $L$ for the hyperbolic cotangent in Eq.~(\ref{eq:def_C(x)}) used to calculate the BCF for the respective $n$ and temperature $T$ (first column). 
The total number $M$ of exponentials used to represent the BCF in Eq.~(\ref{eq:def_EXP_BCF}) is then $M=2 \kappa + L$.
Note that for the combined SD we show only the case $n=5$ in the paper. The other cases are shown in the Supporting Information~\cite{RiEi14_SI_}.
There, also the values of the fit parameters can be found.
}
\end{table}

\section{Conclusions}\label{sec:conclusion}

We have introduced a class of simple functions that are well suited for  approximating many spectral densities (SDs).
A particular feature of our functions is that:

(i) they can exactly describe different algebraic behaviors at small and large frequencies, which allows the treatment of ohmic and superohmic SDs that scale with $\omega^n$ for small $\omega$, where $n$ is odd,

(ii) they lead to a representation of the BCF as a sum of complex exponentials (which is highly desirable for many methods to solve non-Markovian open quantum systems.
The parameters of these exponentials can be calculated analytically.

The number of exponentials (which one wants to keep as small as possible) is twice the number of poles $\omega_{j_k}$ in the upper right part of the complex plane that are needed to represent the SD plus the number of poles used to approximate the hyperbolic cotangent which contains the temperature. 
Due to the form of our functions one typically needs  products with at least $(n-1)/2$ poles $\omega_{j_k}$ for every summand in Eq.~(\ref{eq:_sum_of_anti-symmetrized_terms_scaled}) in order to guarantee that the SD falls off to zero for $\omega\to\infty$

For even and odd scaling parameters $n$ we recognized a crucial difference in the structure of the terms that constitute the BCF.
This we showed exemplarily for the imaginary part of the BCF: 
For odd $n$ different terms than for even $n$ cancel and we arrive at a sum of exponentials, whereas for even $n$ we are left with an expression that indicates that the BCF does not have simple exponential form anymore.

Let us briefly comment on SDs for {\em non harmonic} environments.
For {\em harmonic} environments, as discussed in the present work, knowledge of the BCF Eq.~(\ref{eq:_def_bath_correlation_function}) is sufficient to fully characterize the influence of the environment. In addition the SD $J(\omega)$ is independent of temperature.
For {\em non harmonic} environments one can still define a BCF according to Eq.~(\ref{eq:_def_bath_correlation_function}), however,  one then needs in addition higher correlation functions for the characterization of the influence of the environment. 
Additionally, for non harmonic environments the SD becomes temperature dependent (see e.g.\ Ref.~\cite{We08__}).
The approach of the present paper can still be used to obtain the BCF as a sum of exponentials, however, in general one now has to perform the fitting of the SD for every temperature.

In summary, the presented fit functions form a versatile set of functions, that allow, in combination with the Pad\'e expansion of the hyperbolic cotangent, an efficient representation of bath correlation functions as sums of exponentials.
Since the spectral densities are given in terms of simple poles, they are also well suited for analytical calculations.

\appendix

\section{Open quantum system and absorption}
\label{sec_harm_bath_model}

\subsection{General Hamiltonian}\label{sub_hamiltonian}

Consider a total Hamiltonian of the form
\begin{equation}
\label{H_tot}
  H_{\rm tot}=H_{\rm sys}+H_{\rm int}+H_{\rm env},
\end{equation}
that is, as a sum of the system part $H_{\rm sys}$, the bath $H_{\rm env}$, and the interaction $H_{\rm int}$ between system and bath.
The bath
\begin{equation}
\label{H_bath}
  H_{\rm env}=\sum_{\lambda}\omega_{\lambda}a_{\lambda}^{\dagger}a_{\lambda}
\end{equation}
is assumed to be a set of harmonic modes with frequencies $\omega_{\lambda}$ and annihilation (creation) operators $a_{\lambda}$ ($a_{\lambda}^{\dagger}$).
The interaction between system and bath  is taken as the sum of products of bath ($a_{\lambda}, a_{\lambda}^{\dagger}$) and system ($K$) operators,
\begin{equation}
\label{H_int}
  H_{\rm int}=\sum_{\lambda}\left(\kappa_{\lambda}^*a_{\lambda}K^{\dagger}+\kappa_{\lambda}a_{\lambda}^{\dagger}K\right),
\end{equation}
with coupling constants $\kappa_{\lambda}$.
The coupling to the bath can be encoded in the bath spectral density
\begin{equation}
\label{j_of_omega}
  J(\omega) = \pi \sum_{\lambda}|\kappa_{\lambda}|^2\ \delta(\omega-\omega_{\lambda}),
\end{equation}
which is taken to be a continuous function of the frequency $\omega$.

Note that sometimes $j(\omega)=J(\omega)/\omega^2$ is denoted as the SD.
In these two different conventions, the so-called reorganization energy $E_{\lambda}$ and the total Huang-Rhys factor $X$ are defined as
\begin{equation}
 \begin{split}
  E_{\lambda} &= \frac{1}{\pi}\int_0^{\infty} d\omega\, \frac{J(\omega)}{\omega}
               = \frac{1}{\pi}\int_0^{\infty} d\omega\, j(\omega) \omega,           \\
  X           &= \frac{1}{\pi}\int_0^{\infty} d\omega\, \frac{J(\omega)}{\omega^2}
               = \frac{1}{\pi}\int_0^{\infty} d\omega\, j(\omega).
 \end{split}
\end{equation}

\subsection{Absorption}

For absorption we consider a two-level system with ground state $\ket{g}$ and excited state $\ket{e}$ and Hamitonian $H_{\rm sys}=\epsilon_g\ket{g}\bra{g} +\epsilon_e\ket{e}\bra{e}$, where $\epsilon_g$ and $\epsilon_e$ are the energies of the ground and excited state.
For the interaction of the two-level system with the bath we take the coupling operator $K$ in Eq.~(\ref{H_int}) to be $K=\ket{e}\bra{e}$.
This model for a molecule is discussed e.g.\ in Ref.~\cite{RoStWh12_204110_}.
We consider absorption from the ground state via a dipole transition characterized by the transition dipole operator $\hat{\mu}=\mu\left(\ket{g}\bra{e}+\ket{e}\bra{g}\right)$.

The total initial state of system and bath prior to light absorption is taken as
\begin{equation}
\label{eq:rho_0}
 \rho_0=\ket{g}\bra{g}\otimes \rho_T,
\end{equation}
where the system is in its ground state $\ket{g}$ and the bath is in a thermal state at temperature $T$, i.e.\
$\rho_T = e^{- H_{\rm env}/T} / \mathcal{Z}$ with $\mathcal{Z} = {\rm Tr}_{\rm env}\{ e^{-H_{\rm env}/T}\}.$
Here  ${\rm Tr}_{\rm env}$ denotes the trace over the bath degrees of freedom.
The transition strength for linear optical spectra can be obtained from the half-sided Fourier transform
\begin{equation}
\label{eq:spectrum}
 A(\omega) = {\rm Re} \; \int_0^{\infty} dt \, e^{i \omega t}\, c(t)
\end{equation}
of the dipole correlation function
\begin{equation}
c(t) = {\rm Tr} \{ \hat{\mu}(t)\hat{\mu}(0) \rho_T \}.
\end{equation}

\section{Absorption Spectrum for $T=0$ and Weak System-Bath Coupling}
\label{sec:T0_spectra}

For the temperature $T=0$ the lineshape function $g(t)$ Eq.~(\ref{eq:g(t)_SD}) reduces to
\begin{equation}
\label{eq:g(t)_T0}
 g(t)=-\frac{1}{\pi}\int_{0}^{\infty}d\omega\,\frac{J(\omega)}{\omega^{2}}(e^{-i\omega t}-1),
\end{equation}
where we neglected the summand $-i E_\lambda t$, which simply leads to a shift of the spectrum.
We then Taylor expand $e^{-g(t)}$ in Eq.~(\ref{eq:abs_FT}) and obtain
\begin{equation}
\label{eq:g(t)_T0_Taylor}
 e^{-g(t)} = \sum_{k=0}^{\infty}\frac{1}{k!}\Bigl(\frac{1}{\pi}\int_{0}^{\infty}d\omega\,\frac{J(\omega)}{\omega^{2}}(e^{-i\omega t}-1)\Bigr)^{k}.
\end{equation}
For weak system-bath coupling $J(\omega)$ has small magnitude and we therefore truncate Eq.~(\ref{eq:g(t)_T0_Taylor}) after $k=1$ resulting in
\begin{equation}
 e^{-g(t)}=1+\frac{1}{\pi}\int_{0}^{\infty}d\omega\,\frac{J(\omega)}{\omega^{2}}(e^{-i\omega t}-1).
\end{equation}
For the absorption spectrum Eq.~(\ref{eq:abs_FT}) we find (neglecting the shift $\omega_0$)
\begin{equation}
\label{eq:spec_T0_weak_coupling}
 A(\omega) = (1-X)\delta(\omega)+\frac{1}{\pi}\frac{J(\omega)}{\omega^{2}}
\end{equation}
with $X=\frac{1}{\pi}\int_{0}^{\infty}d\omega\,\frac{J(\omega)}{\omega^{2}}$. We thus see, that the absorption spectrum for $T=0$ and weak system-bath coupling is given by a $\delta$-shaped zero phonon line and a sideband proportional to $J(\omega)/\omega^2$.

\section{Evaluation of the BCF decomposition}\label{sec:evaluation}

\subsection{Application of the Residue theorem \label{sec:_General_Scheme}}

For our fit functions $J(\omega)$ given by Eq.~(\ref{eq:_sum_of_anti-symmetrized_terms_scaled}) the integral Eq.~(\ref{eq:_def_bath_correlation_function}) can be solved analytically using the residue theorem.
One suitable integration contour to apply the residue theorem  is the half-circle over one of the complex half-planes including the real axis (see Fig.~\ref{fig:skizzeInt}).
In order to use this contour, we first extend the integral in Eq.~\eqref{eq:_def_bath_correlation_function}
over the whole real axis.
Here we restrict ourselves to {\em odd}  scaling parameters $n$ as in section  \ref{sec:SD}. The case of {\em even} $n$ is discussed in appendix~\ref{sec:_SD_With_Even_Parameter_n}.

Since we chose $n$ to be an odd number, our fit functions Eq.~(\ref{eq:_sum_of_anti-symmetrized_terms_scaled}) are antisymmetric in $\omega$, i.e.\ they fulfill $J(-\omega) = -J(\omega)$.
Therefore, the full  integrand in Eq.~\eqref{eq:_def_bath_correlation_function} is a symmetric function in $\omega$, because the hyperbolic cotangent and its approximations that we consider are  also anti-symmetric.

We now extend the integration to negative frequencies and write
\begin{equation}
\alpha(t)=a(t)+ib(t)
\end{equation}
with the real part and the respective imaginary part of the correlation
function given by\begin{subequations}\label{eq:_real_and_imaginary_part_of_bcf}
\begin{align}
\begin{split}a(t) & =\frac{1}{2\pi}\int_{-\infty}^{\infty}d\omega\,J(\omega)\coth\left(\frac{\omega}{2T}\right)\cos(\omega t)\\
 & =\frac{1}{2\pi}\int_{-\infty}^{\infty}d\omega\,J(\omega)\coth\left(\frac{\omega}{2T}\right)e^{i\omega t},
\end{split}
\label{eq:_real_part_of_bcf}\\
\nonumber \\
\begin{split}b(t) & =-\frac{1}{2\pi}\int_{-\infty}^{\infty}d\omega\,J(\omega)\sin(\omega t)\\
 & =-\frac{1}{2\pi i}\int_{-\infty}^{\infty}d\omega\,J(\omega)e^{i\omega t}.
\end{split}
\label{eq:_imaginary_part_of_bcf}
\end{align}
\end{subequations}
Here again the symmetry of the integrand was used to introduce exponentials replacing the trigonometric functions. 
These exponentials will simplify the following evaluation of the integrals Eqs.~\eqref{eq:_real_part_of_bcf} and \eqref{eq:_imaginary_part_of_bcf}.

In order to apply the residue theorem to the Eqs.~\eqref{eq:_real_part_of_bcf} and \eqref{eq:_imaginary_part_of_bcf} for $a(t)$ and $b(t)$ we need to know the poles and residues of both our fit functions $J(\omega)$, for $b(t)$, as well as the product $J(\omega)\coth\frac{\omega}{2T}$, for $a(t)$.

Let $\polesA$ denote those poles of  $J(\omega)\coth\frac{\omega}{2T}$ and $\polesB$ those of $J(\omega)$, that lie within the integration contour.
Then one has
\begin{align}
\label{eq:a(t)_poles}
a(t)&= i \sum_{\{\polesA\}} \Res{\polesA}{J(\omega)\coth\frac{\omega}{2T}} e^{i \polesA t}, \\
b(t)&= - \sum_{\{\polesB\}} \Res{\polesB}{J(\omega)} e^{i \polesB t}.
\label{eq:b(t)_poles}
\end{align}
Here we have used that $e^{i \omega t}$ is an analytic function (i.e.\ has no poles) and we used that  $J(\omega)\coth\frac{\omega}{2T}$ and  $J(\omega)$ have only poles of order one~\footnote{Higher order terms lead in general to time-dependent coefficients.}.
Note that Eqs.~(\ref{eq:a(t)_poles}) and (\ref{eq:b(t)_poles}) are of the desired form given by Eq.~(\ref{eq:def_EXP_BCF}).
For our fit functions Eq.~(\ref{eq:_sum_of_anti-symmetrized_terms_scaled}) the residues appearing in Eqs.~(\ref{eq:a(t)_poles}) and~(\ref{eq:b(t)_poles}) can be evaluated analytically.

In order to obtain a finite number of poles for the first expression Eq.~(\ref{eq:a(t)_poles}) we use an expansion of the hyperbolic cotangent given by Eq.~(\ref{eq:def_C(x)}).
Different possibilities for such an expansion are described in appendix~\ref{sec:coth_approx}.

In the following subsections we evaluate  the expressions Eq.~(\ref{eq:a(t)_poles}) and  Eq.~(\ref{eq:b(t)_poles}) for our fit functions Eq.~(\ref{eq:_sum_of_anti-symmetrized_terms_scaled}).
We consider first $b(t)$ and then the slightly more complicated term $a(t)$.

\subsection{Evaluation of $b(t)$}
\label{sub:evaluation_b(t)}

Inserting  the SD Eq.~(\ref{eq:_sum_of_anti-symmetrized_terms_scaled}) into Eq.~(\ref{eq:b(t)_poles}) we see that the relevant poles within the integration contour are $\omega_{j_{k}}$  for $J_k(\omega)$ and $-\omega^*_{j_{k}}$  for $J_k(-\omega)$  (see Fig.~\ref{fig:skizzeInt}).
We abbreviate
\begin{equation}
R_{j_{k}}\equiv \Res{\omega_{j_{k}}}{J_k(\omega)}.
\end{equation}
We find~\footnote{For poles of first order one has \\
$\Res{\omega_{i}}{f(\omega)}=\lim_{\omega\to\omega_{i}} (\omega-\omega_{i})f(\omega)$.} $R_{j_k}=\frac{1}{(\omega_{j_{k}}-\omega^*_{j_{k}})}\prod_{j\ne j_{k}}\frac{1}{(\omega_{j_k}-\omega_{j})(\omega_{j_k}-\omega^{*}_{j})}$, where the product contains all the poles of $J_k$ except $\omega_{j_k}$.
Using the explicit form of $J_k(\omega)$ one can easily calculate that
\begin{equation}
\Res{-\omega^*_{j_{k}}}{J_k(-\omega)}=- R_{j_{k}}^*.
\end{equation}
Thus we can write
\begin{equation}
\begin{split}
b(t)=-\sum_k p_k &\Big(
\sum_{j_k} \omega_{j_{k}}^{n-1} R_{j_{k}} e^{i \omega_{j_{k}} t}\\
&-\sum_{j_k}  (-\omega^*_{j_{k}})^{n-1} (- R_{j_{k}}^*) e^{i(- \omega_{j_{k}}^*) t}
\Big)
\end{split}
\end{equation} 
which is already in the desired form of Eq.~(\ref{eq:def_EXP_BCF}).

Using that $n$ is taken to be odd we can write 
\begin{equation}
b(t)=\sum_k \sum_{j_k} (b_{j_{k}}^{+}  e^{i \omega_{j_{k}} t}
+b_{j_{k}}^{-} e^{-i \omega_{j_{k}}^* t})
\label{eq:_b(t)_final}
\end{equation}
with
\begin{align}
b_{j_{k}}^{+}&=-p_k \omega_{j_{k}}^{n-1} R_{j_{k}}, \\
b_{j_{k}}^{-}&= (b_{j_{k}}^{+})^*,
\end{align}
which makes it also apparent that $b(t)$ is indeed real, as it should be.

Note that the number of exponentials is equal to the number of poles in the upper half-plane used to characterize the SD.
This is $2\times \sum_k \sum_{j_k} 1$.

\subsection{Evaluation of $a(t)$}
\label{sub:evaluation_a(t)}

If the poles $\omega_\ell \equiv 2T \xi_\ell$ of the expansion $C(\omega/2T)$ for the hyperbolic cotangent do not coincide with the poles $\omega_{j_k}$ and $-\omega^*_{j_k}$ of $J(\omega)$, one can write
\begin{equation}
\begin{split}
\label{eq:a(t)_EXP}
a(t)=&i \sum_{\{y\}} \Res{y}{J(\omega)}\, \coth \frac{y}{2T} e^{i yt}\\
& +i\sum_{\ell=1}^L J(w_\ell) \Res{w_\ell}{C\big(\omega/2T\big)} e^{i w_{\ell} t}.
\end{split}
\end{equation}
Here $\{y\}$ denotes those poles of $J(\omega)$ which are inside the integration contour.  
The pole $1/x$ of Eq.~(\ref{eq:def_C(x)}) does not contribute for $n\ge 1$, because for $n\ge 1$ the product $J(\omega)\frac{1}{\omega}$ does not have a pole at $\omega=0$.

Equation (\ref{eq:a(t)_EXP}) is  in the form  Eq.~(\ref{eq:def_EXP_BCF}).
Similar to $b(t)$ we can rewrite this result:
We first note that from
$f(-z^*)=-f(z)^*$ it follows ${\rm Res}_{-z_0^*}\,f=({\rm Res}_{z_0}\,f)^*$.
Because the approximations $C(z)$ that we consider also obey the symmetry $C(-z^*)=-(C(z))^*$ of the exact hyperbolic cotangent, we can write
\begin{equation}
\begin{split}
\label{eq:a(t)_EXP_final}
a(t)=&\sum_k \sum_{j_k} (a_{j_{k}}^{+}  e^{i \omega_{j_{k}} t}
+a_{j_{k}}^{-} e^{-i \omega_{j_{k}}^* t})\\
&+i \sum_{\ell=1}^L J(w_\ell) \Res{w_\ell}{C\big(\omega/2T\big)} e^{i w_{\ell} t} 
\end{split}
\end{equation}
with
\begin{align}
\label{eq:a_{j_k}^+}
a_{j_{k}}^{+}&= -i b_{j_{k}}^{+}\cdot \coth \frac{\omega_{j_k}}{2T} \\
a_{j_{k}}^{-}&= (a_{j_{k}}^{+})^*.
\end{align}
For the approximations (\ref{eq:def_C(x)}) of the hyperbolic cotangent that are typically used the $\omega_{\ell}$ are either purely imaginary or symmetric with respect to the imaginary axis. For purely imaginary $\omega_{\ell}$ the residues are real, also the exponent is real and $J(w_\ell)$ is purely imaginary.
One then sees explicitly that $a(t)$ is a real function. Similar arguments hold for symmetric poles.

\section{Approximating the Hyperbolic Cotangent}
\label{sec:coth_approx}

\subsection{Low-frequency approximations}
Several schemes have been proposed to obtain a suitable pole decomposition of the hyperbolic cotangent.
The most common one is the so-called Matsubara decomposition, where one simply takes a finite number of the exact poles $\nu_\ell=i\pi \ell$ , $\ell=0,\dots,L$ into account.
While being straightforward, this Matsubara decomposition converges very slowly against the exact result so that for low temperatures (i.e.\ large $\omega/2T$) many poles are required.
To overcome this problem more advanced pole decompositions have been suggested.
 For example the {\it partial fraction decomposition} suggested by Croy and Saalmann \cite{CrSa09_073102_} or the {\it Pad\'e decomposition} of  Hu et al. \cite{HuXu10_101106_}. 
The  partial fraction decomposition converges faster than the Matsubara decomposition, but it has an important drawback. 
Because the method is based on finding the roots of a polynomial, which is an ill-conditioned problem in general, one needs high-precision arithmetic to calculate the poles of the approximation correctly. 
Although this does not seem to be a problem for the approximation of the hyperbolic cotangent with real arguments \cite{CrSa10_159904_}, it can lead to severe deviations for the approximation of the integral Eq.~\eqref{eq:_real_part_of_bcf}, when we numerically evaluate it using the residue theorem. 
This is because the spectral density $J(\omega)$ itself contains poles in the complex plane, which then can come very close to the incorrectly evaluated poles of the hyperbolic cotangent leading to large numerical
errors when calculating products of $J(\omega)$ and $\coth(\omega/2T)$.

In order to handle spectral densities with arbitrary poles, it is therefore beneficial to have a pole decomposition of the coth function with only purely imaginary poles (except from the pole at zero). 
Recently such an expansion in simple poles lying on the imaginary axis has been suggested \cite{HuXu10_101106_}. 
It is based on Pad\'e approximants and shows very rapid convergence.
We suggest using this approach for approximating the hyperbolic cotangent.
For completeness we state the final result of Ref.~\cite{HuXu10_101106_} in the form appropriate for our problem.
Following \cite{HuXu10_101106_} one obtains  the parameters $\eta_{j}$ and $\xi_{j}$ occuring in the expansion  Eq.~(\ref{eq:def_C(x)}) by the following procedure. 
Consider the real symmetric matrices $\Lambda$ and $\tilde{\Lambda}$ whose elements are defined as
\begin{equation}
\Lambda_{jk}=\frac{\delta_{j,k\pm1}}{\sqrt{(2j+1)(2k+1)}}\qquad j,k=1,\ldots,2L,
\end{equation}
\begin{equation}
\tilde{\Lambda}_{jk}=\frac{\delta_{j,k\pm1}}{\sqrt{(2j+3)(2k+3)}}\qquad j,k=1,\ldots,2 L -1.
\end{equation}
Then the $\xi_{j}$ and auxiliary parameters $\zeta_{j}$ are given
by 
\begin{equation}
\xi_{j}=\frac{i}{\operatorname{EV}_+[\Lambda]},\qquad\zeta_{j}=\frac{1}{\operatorname{EV}_+[\tilde{\Lambda}]}
\end{equation}
with the symbol $\operatorname{EV}_+[A]$ denoting the positive eigenvalues
of matrix $A$. The parameters $\eta_{j}$ are finally obtained as
\[
\eta_{j}=\frac{1}{2}L(2L+3)\frac{\prod_{i}(\zeta_{i}^{2} + \xi_{j}^{2})}{\prod_{i}(-\xi_{i}^{2} + \xi_{j}^{2})}.
\]

\subsection{Other approximations}
Besides the approximations of the previous section there are other schemes attempting to approximate the hyperbolic cotangent also at high frequencies.
This is typically done by using a high frequency expansion of the hyperbolic cotangent of the form
\begin{equation}
\label{eq:coth_high_frequency_approx}
\coth(x)\approx 1 + \sum^L_{\ell=1} 2\exp(-2 \ell x) + r(x),
\end{equation}
where $r(x)$ is typically a function that tries to take low frequencies into account.

\subsection{Comparison of different approximations of the hyperbolic cotangent}

\begin{figure}
\includegraphics[width=\columnwidth]{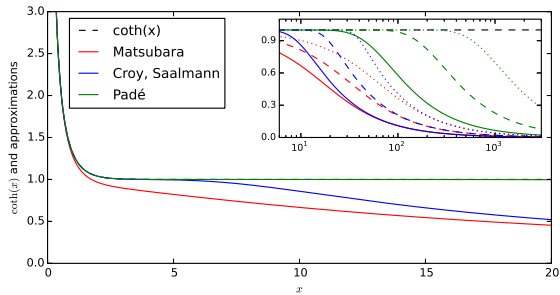}
\caption{\label{fig:_coth_approximations_01}
Main plot: Comparison between $\coth(x)$ (black, dashed), Matsubara (red), Croy/Saalmann (blue) and Padé expansion (green), all with $L=5$ expansion terms (poles with positive imaginary part).
In the displayed $x$-range the Padé approximation and the exact $\coth(x)$ are indistinguishable.
The inset shows the same curves on a logarithmic scale and in addition the cases of $L=10$ (dashed) and $L=20$ (dotted) expansion terms.
}
\end{figure}

\begin{figure}
\includegraphics[width=\columnwidth]{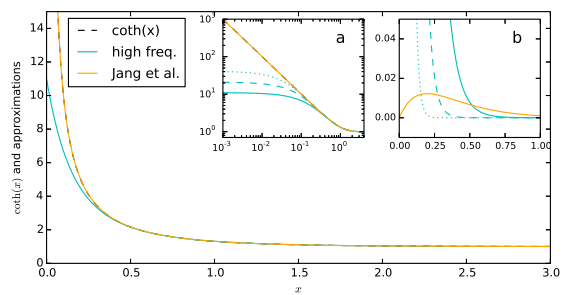}
\caption{\label{fig:_coth_approximations_02}
Main plot: Comparison between $\coth(x)$ (black, dashed) with the high-frequency (low-temperature) expansion  with $r(x) \equiv 0$ (cyan) and the approximation used by Jang et al.\ in Ref.~\cite{JaCaSi02_8313_} (orange). For the high-frequency approximation with $r(x) \equiv 0$ we use $L=5$ expansion terms. In the displayed $x$-range the approximation of Jang et al.\ is indistinguishable from the exact hyperbolic cotangent.
Inset (a) shows the curves from the main plot on a double-logarithmic scale and in addition the cases $L=10$ (dashed) and $L=20$ (dotted) for the high-frequency approximation with $r(x) \equiv 0$.
Inset (b) shows the absolute value of the difference between the approximations of inset (a) and the exact hyperbolic cotangent.
}
\end{figure}

\begin{figure}
\includegraphics[width=\columnwidth]{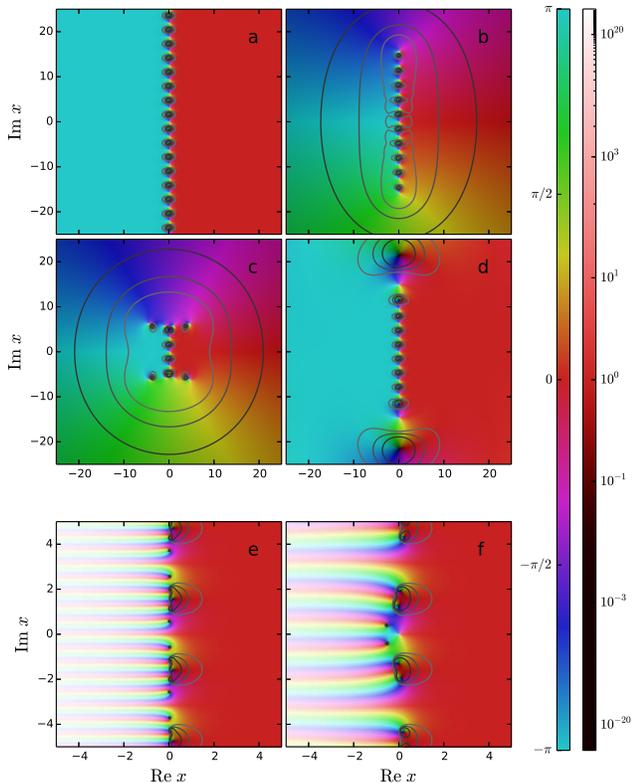}
\caption{\label{fig:_coth_complex}
Panels (a)-(d): Hyperbolic cotangent (a), Matsubara (b), Croy/Saalmann (c) and Pad\'e approximation (d) with $L=5$ for complex arguments $x$.
Panels (e)-(f): high-frequency expansion with $L=5$ and $r(x) \equiv 0$ (e) and the approximation of Jang et al.\ (f) for complex arguments $x$.
The phase is color-coded (compare to left colorbar) and contour lines indicate absolute values corresponding to 0.9 (light gray), 0.7 (medium gray) and 0.5 (dark gray).
The absolute value of the functions is additionaly shown as the lightness of the colors (compare to right colorbar).
Note that the high-temperature expansions (a)-(d) and the two other approximations are displayed on a different scale.
}
\end{figure}

For completeness we discuss briefly some properties of the hyperbolic cotangent and its approximations.

\subsubsection{Low-frequency approximations}

In Fig.~\ref{fig:_coth_approximations_01} we show a comparison of the exact hyperbolic cotangent (black, dashed) with the Matsubara (red), the Croy/Saalmann (blue) and the Padé expansion (green), all with $L=5$ expansion terms (poles with positive imaginary part).
In the displayed $x$-range the Padé approximation and the exact $\coth(x)$ function are indistinguishable. 
The inset shows the same curves on a logarithmic scale and in addition the curves for $L=10$ (dashed) and $L=20$ (dotted) expansion terms.
One clearly sees the superior behavior of the Padé approximation.

In Fig.~\ref{fig:_coth_complex} the hyperbolic cotangent and its approximations are shown in the complex plane.
In  subplot (a) the exact hyperbolic cotangent is shown. 
In the subplots  (b), (c), (d) we show the Matsubara, Croy/Saalmann and Padé expansion for the case of  $L=5$ terms (as in Fig.~\ref{fig:_coth_approximations_01}).
The phase is color-coded and the absolute value is indicated as lightness and additionally with contour lines.
The intersection of these four plots with the (positive) real axis corresponds to the curves displayed in Fig.~\ref{fig:_coth_approximations_01}.
In Fig.~\ref{fig:_coth_complex}~(b)-(d) one sees that all three approximations also obey the anti-symmetry $\coth(-x)=-\coth(x)$ of the exact hyperbolic cotangent, which we have used in Eqs.~(\ref{eq:_real_part_of_bcf}) and (\ref{eq:_imaginary_part_of_bcf}), to extend the integral to negative arguments. 
However, they have quite different pole structures.
One clearly sees that the Pad\'e approximation is not only superior on the real axis, but approximates the hyperbolic cotangent much better than the other expansions in the complex plane.

Note that in the case of the Croy/Saalmann decomposition several poles have real parts different from zero.
Thus one has to be careful that these poles do not coincide with or come close to the poles of the fit function $J(\omega)$.

\subsubsection{High-frequency-type approximations}
\label{sub:coth_high_frequency_approx}

As already noted in section \ref{sec:coth} a particular appealing approximation of the hyperbolic cotangent has been suggested by Jang et al. \cite{JaCaSi02_8313_}, where $L=2$ and $r(x)=\exp(-5x)/x$ in Eq.~(\ref{eq:coth_high_frequency_approx}). It thus possesses only 4 simple terms and is able to approximate the hyperbolic cotangent very well on the real axis.
This is shown in Fig.~\ref{fig:_coth_approximations_02}, where the  approximation of Jang et al.\ from Ref.~\cite{JaCaSi02_8313_} (orange) is shown together with the exact hyperbolic cotangent (black, dashed). 
In addition, we show results of a simple high frequency expansion with $r(x) \equiv 0$ for $L=5$ expansion terms.
One sees that the approximation of Jang et al.\ is nearly perfect over the whole range of arguments.
The high frequency expansion with $r(x) \equiv 0$ also does well for larger arguments, but it clearly deviates from the exact hyperbolic cotangent for smaller arguments.

However, both functional forms, the approximation of Jang et al.\ as well as the high frequency expansion with $r(x) \equiv 0$, are expansions around \emph{positive} infinity and diverge for arguments with negative real part (see Fig.~\ref{fig:_coth_complex}, panels (e), (f)). Thus, they do not preserve the anti-symmetry of the hyperbolic cotangent so that the extension of the integral domain as in Eq.~(\ref{eq:_real_part_of_bcf}) fails and the residue theorem cannot be used anymore \footnote{We were also not able to anti-symmetrize Eq.~(\ref{eq:coth_high_frequency_approx}) such that it is still meromorphic.}.

We tried to solve the integral Eq.~(\ref{eq:_def_bath_correlation_function}) with our fit functions Eq.~(\ref{eq:_sum_of_anti-symmetrized_terms_scaled}) and the approximation Eq.~(\ref{eq:coth_high_frequency_approx}) for the hyperbolic cotangent, but we only succeeded in obtaining the exponential form Eq.~(\ref{eq:def_EXP_BCF}) for $L=0$ and $r(x) \equiv 0$. 
The reason for this seems to be that the exponentials in Eq.~(\ref{eq:coth_high_frequency_approx}) contain arbitrary integer powers of $x$ (i.e.\ even \emph{and} odd powers).
 The problems that arise in this case can be seen from the integrals in appendix~\ref{sec:_SD_With_Even_Parameter_n}.

\section{Spectral Densities with Even Parameter $n$
\label{sec:_SD_With_Even_Parameter_n}}

In section~\ref{sec:eval_BCF} we considered
only fit functions that behave like $\omega^n$ for small frequencies, where $n$ is an \emph{odd} positive integer.

The reason for this restriction was the following:
Our method of evaluation relied on the use of the residue theorem and the integration contour shown in Fig.~\ref{fig:skizzeInt}.
To this end we made use of the anti-symmetry of our fit functions in $\omega$ (for $n$ odd).
For even $n$ Eq.~(\ref{eq:_sum_of_anti-symmetrized_terms_scaled}) is a \emph{symmetric} function in $\omega$ and an anti-symmetrization thereof  results in a non-meromorphic function, where the residue theorem cannot be applied.

In the following, we approach the question whether one can find a similar decomposition of the bath correlation function in terms of exponential functions also for even $n$ \footnote{At the moment we can make no statement for $n$ not belonging to the set of natural numbers.}.

We now solve the integral Eq.~\eqref{eq:_def_bath_correlation_function} using Fourier transformations. 
To keep the analysis as simple as possible we now consider a fit function with only one contribution $J_{k}(\omega)$ in Eq.~\eqref{eq:_sum_of_anti-symmetrized_terms_scaled} and one (first-order) pole $\omega_{j_k}$ ($k=j=1$)
in Eq.~\eqref{eq:_pole_decomposition} with
\[
\omega_{\rm p} \equiv \omega_{1_1} = \Omega+i\gamma,
\]
where $\Omega$ and $\gamma$ are both positive. (This structure of
$J_{k}(\omega)$ is the one that was used in Eq.~\eqref{eq:_lorentzian} in section~\ref{sec:examples}.)
Thus the total fit function we now consider has the form
\begin{equation}
J(\omega) = \omega^{n-1} p\left(\frac{1}{(\omega-\omega_{\rm p})(\omega-\omega_{\rm p}^{*})}-\frac{1}{(\omega+\omega_{\rm p})(\omega+\omega_{\rm p}^{*})}\right).\label{eq:_lorentzian_scaled}
\end{equation}
In the following, we consider only the imaginary part $b(t)$, defined in Eq.~\eqref{eq:_imaginary_part_of_bcf}, since it already allows us to see the differences that arise between even and odd scaling parameters $n$. 
Because we evaluate $b(t)$ using Fourier transforms it is convenient to write the sine funcion in exponential form and use a partial fraction decomposition of the SD in Eq.~\eqref{eq:_lorentzian_scaled}. 
One then obtains
\begin{equation}
b(t)=\frac{1}{2\pi i}\frac{p}{\omega_{\rm p}-\omega_{\rm p}^{*}}\left(A_{\omega_{\rm p}}-A_{\omega_{\rm p}^{*}}+A_{-\omega_{\rm p}}-A_{-\omega_{\rm p}^{*}}+\mathrm{c.c.}\right)\label{eq:_b(t)_with_A_z}
\end{equation}
with c.c.\ denoting complex conjugate terms and $A_{z}$ defined through
\begin{equation}
A_{z}=\int_{0}^{\infty}d\omega\,\omega^{n-1}\frac{1}{\omega-z}e^{-i\omega t}.\label{eq:_def_A_z}
\end{equation}
These integrals Eq.~\eqref{eq:_def_A_z} are now extended to negative frequencies by introducing the Heaviside
step function $\Theta(\omega)$.
Replacing the multiplications with $\omega^{n-1}$ by derivatives with respect to time gives
\begin{equation}
A_{z}=\Bigl(i\frac{\partial}{\partial t}\Bigr)^{n-1}\int_{-\infty}^{\infty}d\omega\,\Theta(\omega)\,\frac{1}{\omega-z}e^{-i\omega t}.
\end{equation}
Using the convolution theorem we find
\begin{equation}
\begin{split}A_{z}=\Bigl(i\frac{\partial}{\partial t}\Bigr)^{n-1}\int_{-\infty}^{\infty}dt'\,\biggl( & \int_{-\infty}^{\infty}d\omega\,\Theta(\omega)\, e^{-i\omega(t-t')}\\
 & \times\frac{1}{2\pi}\int_{-\infty}^{\infty}d\omega'\,\frac{1}{\omega'-z}e^{-i\omega't'}\biggr).
\end{split}
\end{equation}
After inserting the known Fourier transforms of $\Theta(\omega)$ and $1/(\omega-z)$ we obtain
\begin{equation}
\begin{alignedat}{4}
A_{z}
 & = \Bigl(i\frac{\partial}{\partial t}\Bigr)^{n-1}\int_{-\infty}^{0}dt' & & f(t-t')
 \begin{cases}
+ie^{-izt'} & :\ \mathrm{Im}[z]>0\\
0 & :\ \mathrm{otherwise}
 \end{cases}\\
 & + \Bigl(i\frac{\partial}{\partial t}\Bigr)^{n-1}\int_{0}^{\infty}dt' & & f(t-t')
 \begin{cases}
-ie^{-izt'} & :\ \mathrm{Im}[z]<0\\
0 & :\ \mathrm{otherwise}
 \end{cases}
\end{alignedat}
\label{eq:_A_z_final}
\end{equation}
with $f(\tau) = -\frac{i}{\tau} + \pi\delta(\tau)$.

Writing out all eight contributions in Eq.~\eqref{eq:_b(t)_with_A_z} with Eq.~\eqref{eq:_A_z_final} one sees that for odd numbers $n$ the terms under the integral proportional to $1/(t-t')$ exactly cancel and only
the $\delta$-terms survive. Thus the bath correlation function $b(t)$
for odd $n$ is
\begin{equation}
b(t)\overset{n\textrm{ odd}}{=}-\frac{p}{\omega_{\rm p}-\omega_{\rm p}^{*}}\Bigl(i\frac{\partial}{\partial t}\Bigr)^{n-1}(e^{i\Omega t}-e^{-i\Omega t})e^{-\gamma\left|t\right|},
\end{equation}
which is --- after calculating the contained derivative --- a (complex)
exponential function in $t$ and identical (for $t>0$) to the result Eq.~\eqref{eq:_b(t)_final} obtained from the residue theorem. For even numbers
$n$ now the $\delta$-terms in the $A_{z}$ cancel in Eq.~\eqref{eq:_b(t)_with_A_z} and we are left with
\begin{equation}
\begin{split}b(t)\overset{n\textrm{ even}}{=} & -\frac{1}{\pi i}\frac{p}{\omega_{\rm p}-\omega_{\rm p}^{*}}\Bigl(i\frac{\partial}{\partial t}\Bigr)^{n-1} \\
 & \times\int_{-\infty}^{\infty}dt'\,\frac{1}{t-t'}(e^{i\Omega t'}-e^{-i\Omega t'})e^{-\gamma\left|t'\right|}.
\end{split}
\end{equation}
This integral written in terms of special functions reads
\begin{equation}
\begin{alignedat}{2}b(t)\overset{n\textrm{ even},\ t > 0}{=}\frac{1}{2\pi i}\frac{p}{\omega_{\rm p}-\omega_{\rm p}^{*}}\cdot2\Bigl(i\frac{\partial}{\partial t}\Bigr)^{n-1} \\
 & \hspace{-4.5cm}\times\biggl[ & \Bigl( & e^{i\omega_{\rm p}t}\operatorname{Ei}(-i\omega_{\rm p}t) -
  e^{-i\omega_{\rm p}^* t}\operatorname{Ei}(i\omega_{\rm p}^* t)\Bigr) \\
 & \hspace{-4.5cm} & + \Bigl( & e^{i\omega_{\rm p}^* t}\operatorname{\Gamma}(0, i\omega_{\rm p}^* t) -
  e^{-i\omega_{\rm p} t}\operatorname{\Gamma}(0, -i\omega_{\rm p} t)\Bigr)\biggr].
\end{alignedat}
\end{equation}
For the exponential integral $\operatorname{Ei}(z)$ and the incomplete gamma function $\operatorname{\Gamma}(a, z)$ we use the definitions according to Abramowitz and Stegun \cite{AbSt64__}.
This result indicates that for even $n$ the imaginary part $b(t)$ of the correlation function does not have simple exponential form anymore.


\end{document}